\documentclass[runningheads,a4paper]{llncs}
\usepackage{amsmath}
\usepackage{amssymb}
\usepackage[linesnumbered,ruled,vlined]{algorithm2e}
\usepackage{float}
\usepackage{xfrac}
\usepackage[all]{xy}

\usepackage{multirow}
\usepackage{dirtytalk}
\usepackage{booktabs}
\usepackage{graphicx}
\begin{document}

\title{Ambient Assisted Living technologies from the perspectives of older people and professionals}
\titlerunning{AAL technologies from the perspectives of older people and professionals}  
%
\author{Deepika Singh\inst{1}, Johannes Kropf\inst{1}, Sten Hanke\inst{1} \and Andreas Holzinger\inst{2} }
\authorrunning{Deepika Singh et al.} 
%
\tocauthor{Deepika Singh, Johannes Kropf, Sten Hanke, and Andreas Holzinger}
\institute{AIT Austrian Institute of Technology, Austria\\ 
\email{deepika.singh@ait.ac.at}\\
\and
Holzinger Group, HCI-KDD, Institute for Medical Informatics/Statistics,\\
Medical University Graz, Austria}

\maketitle              

\begin{abstract}
Ambient Assisted Living (AAL) and Ambient Intelligence technologies are providing support to older people in living an independent and confident life by developing innovative ICT-based products, services, and systems. Despite significant advancement in AAL technologies and smart systems, they have still not found the way into the nursing home of the older people. The reasons are manifold. On one hand, the development of such systems lack in addressing the requirements of the older people and caregivers of the organization and the other is the unwillingness of the older people to make use of assistive systems. A qualitative study was performed at a nursing home to understand the needs and requirements of the residents and caregivers and their perspectives about the existing AAL technologies.
\keywords{Smart homes, ambient assisted living, independent living, ageing, quality of life}
\end{abstract}
\section{Introduction}
The aging population is the one of the major concerns of the world due to its direct socioeconomic implications. According to the United Nations’ report, the number of people aged 65 years and above is expected to grow from 901 million in 2015 to 1.4 billion in 2030 \cite{WPA2015,kleinberger2007ambient}. According to the European population projections, it is expected that by 2040, one third of the elderly population will be aged 80 years and above. This demographic changes in the population is due to declines in fertility rate, continuous increases in the life expectancy and the retirement of the baby-boom generation \cite{WPA2015,olderpeople}.\\

Over aging leads to many problems ranging from basic functional disabilities to severe health problems e.g. osteoarthritis, diabetes, depression, chronic obstructive pulmonary diseases and dementia. In addition to the medical problems, the fact of being dependent on family members and/or care providers for their daily activities, cause embarrassment, social inactivity and poor nutrition etc. For long-term care and continuous assessment of physical and mental health in the older people there has been an increasing demand of nursing homes in the last decade \cite{longtermcare,ribbe1997nursing}. However, this does not solve their problems completely; thus technology is the tool which can provide them an independent and happier life and at the same time accurate and timely personal care by the nursing home staff.\\

To overcome some of the mentioned problems, there has been a rapid development in ambient assisted living technologies (AAL) in Europe \cite{blackman2016ambient}. Different AAL solutions towards home monitoring, fall detection, social interaction have been developed using machine learning techniques. Technologies such as smart homes, assistive robots, mobile and wearable sensors have gained a lot of attention, but there are still many challenges that need to be addressed \cite{rashidi2013survey}. The concept of the smart home is a promising and cost-effective way of improving home care for the older people and the disabled in a non-obtrusive way, allowing greater independence, maintaining good health and preventing social isolation. There have been major advancements in the area of smart homes research that is enriching the home environment with technology (e.g., sensors and interconnected devices in a network) \cite{chan2009smart}. The most popular smart home projects are MavHome (Managing an Adaptive Versatile Home) \cite{das2002role}, GatorTech \cite{helal2005gator}, CASAS smart home \cite{rashidi2008keeping}, EasyLiving project \cite{krumm2000multi} etc.  The design of smart home depends on user requirement and living lifestyle of the resident; several approaches are proposed to identify activities of daily living of the resident of the home \cite{skubic2009smart}. Although, the detection of a specific activity depends on the selection of appropriate set of sensors, data processing techniques and effective algorithms to understand daily lifestyle and human behavior \cite{ni2015elderly,skubic2009smart}. The development of such systems lack in addressing the requirements of the older people and caregivers of the organization. One possibility to overcome this issue is proposed by applying extreme usability methods \cite{holzinger2005extreme}. Additionally, various machine learning approaches have been used to develop an intelligent system for activity recognition; most common methods are Nave Bayes \cite{tapia2004activity}, Hidden Markov Model \cite{nguyen2005learning}, and Conditional Random Field \cite{nazerfard2010conditional} classifiers. Despite the number of machine learning approaches, the accuracy of activity recognition is still not robust and unable to deal with uncertainty. There exist lot of challenges that need to be addressed in implementing effective solutions for older people \cite{sun2009promises}. \\

To develop adaptable smart home technologies according to the specific needs of the older residents, their suggestions and inputs are much needed. A number of studies have been carried out where older people are participating in providing suggestions and opinions about the technologies \cite{dongen2016successful}. Residents and family members indicated that they feel safe and had an overall positive attitude towards devices and installation of sensors \cite{demiris2004older} knowing that someone is monitoring them for their wellbeing \cite{alam2012review}. The parameters of quality of life are usually evaluated by the older people on the basis of social contacts, dependency, health, material circumstances and social comparisons \cite{netuveli2008quality}. Certainly, AAL solutions and assistive technologies have a positive impact on different dimensions of health and quality of life. The needs and problems of older people can be addressed by applying appropriate solutions which influence the physical, mental and social dimensions of quality of life \cite{siegel2014contributions}. Nevertheless, the  most important aspect is the acceptance of technology/devices by the older people and professionals \cite{holzinger2008investigation}. \\

Within the framework of this study, we wanted to know the different perspectives of the older people and care giving professionals of a particular nursing home. From the older people  point of view we were interested in the following things: i) how comfortable they are with the existing technologies (sensors, cameras, robots); any ii) specific need they would require assistance or iii) activities which they could imagine to be monitored; iv) their problems and fears. From the caregiver point of view we were interested in the following: which data/information would be useful for them to provide accurate and timely attention to the patients. The study is aimed to understand the needs better so that more useful AAL solutions can be developed considering the inputs from the residents and caregivers. 

\section{Methodology}

\subsection{Planning}
The research activity was planned at Zuyderland nursing home situated in Sittard-Geleen (The Netherlands) for $4$ weeks. This nursing home is a part of Zuyderland Medical and Health care Group in the province of Limburg, Netherlands. The nursing home has a total of $273$ residents and it is divided into small scale living apartments ($48$ residents), elderly care apartments ($100$ residents) and $3$ apartment’s blocks for independent living ($125$ residents). The residents from the elderly care apartments and independent living apartment were contacted to participate in the study.\\ 

The research activities to be performed were divided for each week such that in the first week we collected information about the facilities which are provided to the patients in the hospital and the nursing home. The interviews with the residents and professionals of the hospital and the nursing home were conducted in the second week. The next two weeks were spent on transcribing the answers of the interview questions and analysis of the data obtained from the interviews. The aim of the study was to acquire knowledge about the daily living lifestyle of the old people and to investigate the needs, problems they face. In addition, it was also beneficial to know from the opinions of residents and caring staff about various existing ICT solutions and ambient assisted technologies including Smart Care Home. The caregivers had prior knowledge of existing AAL technologies from former research projects.

\subsection{Focus Group and Sampling}
Care professionals and residents of the nursing home were contacted to ask for the participation in the study. The aim was to enroll large number of people who can best discuss and share their experiences. The residents for the study were chosen by the nursing home’s caring staff since they were knowing their exact medical conditions. For example, this activity should not to disturb any of the medical routines of residents, therefore, the residents suffering from severe chronic diseases (such as cancer and last stage heart diseases), and those with pacemakers and other required monitoring electronic devices such as ECG, pulse oximetry etc. have been excluded from the study. The focus group of the study was composed of residents and professionals as specified below: \\
$1$. Residents from the independent living apartments\\
$2$. Residents from elderly care apartments\\
$3$. Professionals include Physiotherapist, Occupational therapist and Dietician\\ 
$4$. Caring staff of the nursing home\\
$5$. Innovation and development experts \\ 

The total sample size was consisting of seven residents ($n=7$) and six professionals ($n=6$). The characteristics of all the $13$ participants interviewed are shown in Table \ref{residents} and Table \ref{professionals}. We have assigned code to each participant such that \textit{\say{R}} and \textit{\say{P}} denote resident and professional respectively. The average age of the residents was $80$ years and the average experience of the professional in dealing with problems of older people was $17$ years.\\
\begin{table}[tbh]
\centering
\caption{Demographics of the residents}
\setlength{\tabcolsep}{6pt}
\begin{tabular}{cccc}
	\hline
ID& Age (years)  &  Gender  &  Work Status \\ 
	\hline
R$1$&	$90$	&	Male   &	\multirow{7}{*}{\parbox{2cm}{All of them 
were retired}} \\
R$2$& 	$82$	&   Male   &     \\
R$3$& 	$80$	&	Female &     \\
R$4$&   $74$	&	Female &     \\
R$5$&	$65$  	& 	Male   &     \\
R$6$&	$86$	&   Female &     \\
R$7$& 	$81$ 	& 	Female &     \\

	\hline
\end{tabular}
\label{residents}
\end{table}
\begin{table}[tbh]
\centering
\caption{Details of professionals}
\setlength{\tabcolsep}{6pt}
\begin{tabular}{ccccc}
	\hline
ID& Age (years)  &  Gender  &  Specialization	&  Experience with elderly(in years) \\ 
	\hline
P$1$&	$53$	&	Female	&	Dietician				& $20$ \\	 
P$2$& 	$39$	&   Male	&   Physiotherapist			& $16$\\
P$3$& 	$33$	&	Female	&   Occupational therapist	& $11$ \\
P$4$&   $37$	&	Male 	&   Caregiver				& $21$\\
P$5$&	$39$  	& 	Male	&   Innovation expert		& $17$\\
P$6$&	$40$	&   Male	&   Innovation expert		& $18$\\

	\hline
\end{tabular}
\label{professionals}
\end{table}

\subsection{Data Collection}
The interview questions for the residents were finalized after having discussions with care home professionals ranging from nurses (who are in direct contact with the residents) to the innovation experts and vice-versa. The inputs from them were well addressed in the final questionnaires in both the cases. As already stated, the final aim of the interviews was to answer the following questions:
\begin{itemize}
\item[$\bullet$] What are the problems faced by the elderly in performing the daily living activities?
\item[$\bullet$] How smart care homes can contribute to the needs and requirements of the elderly and caregivers?
\item[$\bullet$] An opinion about existing AAL technologies and how it can be improved?
\end{itemize}

In the beginning of the interview, the objective of the study were explained to the participants and they were told that they can leave or interrupt the interview whenever they feel any discomfort and/or unwillingness to answer. Since the interview were conducted in an informal way of talking, the timing was varied from $25$ to $60$ minutes. Consents were taken for audio recording the sessions so as not to miss any important points and actual responses can be quoted. All the participants agreed to audio record the interviews and the data collected were transcribed verbatim for the accuracy immediately after the interview.\\

The questions were framed in the simplest way for sake of clarity; they were also allowed to seek any sort of clarification needed. In order to make the residents comfortable, the interviewer started with an informal conversation, e.g.\\ 
\textit{Q: What are your hobbies?}\\
\textit{Q: What do you like to do in your free time inside the home?}\\ 
And the interview progressed in a manner of fluent conversation. However, the framed questions were asked clearly in the running conversation to be consistent with the responses so that it may not lead to any ambiguity in comparison. \\

The interviews with the residents were performed in two groups, one group was the residents from independent living apartments \textit{(R1-R4)} and other group was of residents from the elderly care apartments \textit{(R5-R7)}. Interviews with professionals and caregivers were conducted on one-to-one basis.
        
\subsection{Ethics }
The research was conducted with the prior permission and after the approval from the hospital and care organization authorities. To ensure confidentiality, an agreement was signed by the researcher with the organization. Full anonymity was promised to all the participants.
\subsection{Data Analysis}
The transcribed data was matched with the audio taped version to remove any discrepancies. Data analysis was performed using qualitative content analysis approach \cite{elo2008qualitative,miles1994qualitative}. Analysis process includes various steps:
\begin{itemize}
\item[$\bullet$]Organizing and collection of data;  
\item[$\bullet$]repeated reading of the data; 
\item[$\bullet$]look for meaningful data and labeling it into codes;  
\item[$\bullet$]grouping the codes in subcategories; 
\item[$\bullet$]grouping subcategories into categories to generate themes.
\end{itemize}

To ensure the rigor in this study following criteria have been fulfilled: i) getting familiar with the residents to form trusting relationships, verifying responses with the participants (credibility); ii) selection of the participants (dependability); iii) using extracts from the interviews to support findings (transferability); iv) and establishing an audit trail (confirmability) \cite{krefting1991rigor}.

\section{Findings and Discussions}
With consideration to the care needs, we decided to perform a thematic analysis focusing on different important aspects of healthy living; three themes were identified: \textit{daily living, social engagement and technology}. The themes have different categories and subcategories, as shown in Table \ref{care}. We have also included some quotes from residents and professionals in the discussion section to highlight the exact needs without generalization.
\begin{table}[tbh]
\centering
\caption{Care needs of the elderly}
\setlength{\tabcolsep}{6pt}
\begin{tabular}{ccc}
	\hline
Themes& Categories  &  Subcategories \\ 
	\hline
\multirow{7}{*}{Daily Living}& \multirow{5}{*}{Problems and needs}	&	Eating  \\
 
& 	&  Cooking       \\
& 	&  Grooming     \\
&	&  Housekeeping    \\
&	&  Bathroom usage    \\\cmidrule{2-3}
& 	\multirow{2}{*}{Psychological care}	& 	Privacy   \\
& 	& 	 	Safety     \\
\hline
\multirow{6}{*}{Social Engagement}& \multirow{3}{*}{Participation in social activities}	&	Social events in elderly home  \\
&	&	   Special events      \\
& 	& 		Visit with family members     \\\cmidrule{2-3}
& 	\multirow{3}{*}{Physical activities}	& 	Exercise   \\
& 	& 	 	Indoor activities     \\
&	&	   Outdoor activities     \\
\hline
\multirow{3}{*}{Technology}& \multirow{3}{*}{Adaptability with technology} & Devices  \\
& 	& 	Robots     \\
& 	& 	Security cameras     \\
	\hline
\end{tabular}
\label{care}
\end{table}

\subsection{Daily living}
In our study we observed the daily living lifestyle of the residents in the nursing home. All the residents have their own private apartments in assistive environment with $24 \times7$ availability of the caring staff.\\

The residents have different problems and needs which are mainly defined by their health conditions. The residents suffering from some physical disabilities, mild chronic diseases and fractures, find difficulty in performing most of their daily activities (such as going to toilet, showering, walking or moving inside their apartment, eating), therefore, they seek more assistance from caregivers. The caregivers help the residents in order to follow their daily routine.\\
\textit{P3: \say{I watch them and analyze that they find difficulties in moving or walking, I look for possibilities so that person can move in their room and actively participate in the environment either with care or some aids.}}\\
\textit{P3: \say{The first thing which the patients want to do without any assistance is going to the toilet; as dependence on care givers brings the sense of embarrassment among the patients.”}}\\

The older people sometimes do not express their urges out of embarrassment which causes poor nutrition and imbalance diet. \\
\textit{R6: \say{I cannot pick something from refrigerator, or from table because I cannot move. Every time I have to ring and call the nurse”}}\\
\textit{P1: \say{I met a patient who told me that I eat less, because I have call somebody to get me something to eat and drink}}\\
In some cases, they need assistance even for small things such as picking up bottle for drinking, cloths from the wardrobe etc.\\
\textit{P2: \say{Residents also need assistance in household activities like closing the door, curtains, turning off the TV and all these activities takes lot of time of the nurses}}\\

From the interview we found out the major factors which affect the patients physiologically are: feeling of embarrassment; helplessness and loneliness; depression and lack of motivation.\\
\textit{R5: \say{My hobbies are painting but I do not have motivation of doing it}}\\
One of main challenges for the caregivers is monitoring the resident who do not interact with anyone and keep themselves inside the apartment. In that case, the caregivers cannot track the patient’s daily activities and prescribed dietary plan. In addition, they sometimes lie and even refuse to take any help by the caregivers. \\
\textit{P1:\say{A patient answered when offered help: Oh No, I don’t need anything. Don’t do that, don’t make anything for me}}

\subsection{Social Engagement and Physical Activities}
Engagement of the residents in social and physical activities depends on the personal choices, interests and their health conditions. From the interviews, we analyzed that most of the residents participate in the social activities organized within the nursing home but only very few of them go out for some physical activity, walking or swimming. They like to go out only with family member and on some special occasions. This unwillingness could be due the fear of fall, fear of exhaustion and thus always need a companion with them.\\

There are few challenges with the residents suffering from severe chronic disease, disabilities; such residents do not like to participate mainly due to the feeling of helplessness and embarrassment. The other major factor which cause less physical activity and less exercise is due to the pain.\\
\textit{P2: \say{Pain has a strong influence on daily living and most important domain where we can act as physical therapist. Would be good if I know in which activities and movement patients suffers pain}}\\

The residents participate in the exercise sessions and trainer motivates them to continue with their exercise. But sometimes exercise during the sessions is not just enough and they are advised to follow a routine. However, it is very difficult to know from the older people that whether they are exercising in their homes or not? \\
\textit{P2: \say{I do not know whether the patient exercise in their own home or not}}\\
The less physical activity also causes poor nutrition; as the patients do not consume food and/or liquid if they remain inactive for a longer duration.\\

From the perspective of professionals, quality of life can be improved by motivating and engaging older people in different social activities of their interest which keep them physically active, thus decrease depression and feeling of loneliness.

\subsection{Technology}
The adaptability with technology is one the most important point to be considered while designing a smart home. It does not solve the purpose if a highly sophisticated device is provided but they do not feel comfortable in using it. We found out from the responses of the older people that they find the technology beneficial and useful especially from the safety point of view; but they have their preferences and restrictions.\\
\textit{P2: \say{When residents see everybody is using it then they use it}}

\begin{table}[tbh]
\centering
\caption{Technology}
\setlength{\tabcolsep}{6pt}
\begin{tabular}{cccc}
	\hline
ID& Smart phone  &  Tablet  &  Laptop/Computer \\ 
	\hline
R$1$&	No	&	Yes	&	No			 \\	 
R$2$& 	No	&   Yes	&   Yes			  \\
R$3$& 	No	&	Yes	&   No	    \\
R$4$&   Yes	&	Yes &   No	    \\
R$5$&	Yes & 	Yes	&   Yes\\
R$6$&	No	&   Yes	&   No\\
R$7$&   No 	& 	No &     Yes\\
	\hline
\end{tabular}
\label{tech}
\end{table}

Table \ref{tech} highlights the comfort levels of the older people in using various devices among smart phone, tablet and laptop/computer. As it is clear from the table, among all the devices, majority of the residents find tablet more comfortable in using than smart phones and computer.\\

In the study, we also inquired from the residents whether they are comfortable with robots and they like to see robots assisting them in daily activities in their home. Most of them disagreed with having robots inside their apartments. Residents pointed out that they are more comfortable in using tablets and agreed in using wearable devices but do not want big robots around them. \\
\textit{R4:\say{No! I do not want robot inside my apartment}}\\
\textit{R1:\say{I cannot cuddle it and they do not give hugs}}\\

In concern with the home security, we asked whether they like to put cameras outside their apartment door to see who visited them, in their absence. The residents strictly denied that they do not want to put camera inside and outside their apartment.\\
\textit{R2: \say{If someone wants to meet me, he/she can come again}}\\
From the perspective of the older people and professionals, a personalized intelligent assistive technologies will be desirable which can provide independence, ensure safety, and should be adaptable according to needs of the residents.
\subsection{Suggestions/Recommendations}
As one of the main motives of the study was to seek suggestions from the residents and professionals for making the nursing home smart. \\
\textit{P1: \say{when I know how active the patient is, it would be nice for me to know how much energy he/she spent per day and will help me in making diet plan and improve accordingly. Now it is always guess and I have to ask patient every time}}\\
\textit{P1: \say{Would be good if there is some system which help in cooking and preparing the meals. In kitchen, if stove get automatically turned off when not in use}}\\
In general, from the caregivers’ perspective, some family members of the residents want to keep a track of their health status and daily updates from the nursing home. A system or application which sends out daily notifications to family members and notify them in case any emergency, would be really desirable. \\
\textit{P4: \say{Would be good to have system for caregiver and resident in the apartment which controls the lighting, door open/close, curtains open/close and inside temperature level}}\\
\textit{P3: \say{If I could know information about small activities from smart home like whether patient is going to bathroom by their own, getting meal or drinks from the kitchen, would be very helpful}}\\
\textit{R3: \say{It would be good if I know my sleeping patterns and activity level inside my home, so I can show that data to my doctor}}\\

From the perspective of the residents safety is the major concern. In the nursing home, residents have a personal alarm system, which they find very useful and beneficial. They use it for calling the caregiver in case of any assistance required and emergency inside the home. However, the residents recommended that such kind of system would be really useful, if they could use it when they are out. In that case, they would feel safer in going out alone, which enhances physical activeness and social involvement.

\section{Conclusion}
The findings of our study provide insights into the problems and needs of the residents of the nursing home. It also highlights the challenges faced by the caregivers during monitoring of the older people. Suggestions and recommendations have been pointed out from both the sides aiming at an independent and confident life of the residents. We summarize them as follows:\\
Although, the needs and problems are varied according to the individual’s needs. The residents from the independent living apartments \textit{(R1-R4)} do not really want home automation as they can perform basic daily activities. They think it can make them less physically active. On the other hand residents from the elderly care apartments \textit{(R5-R7)} get regular assistance by the caregivers and there is evident need of a smart care home to reduce less personal assistance in basic household activities such as lock/unlock doors, opening/closing window shields, lights controlling etc. However, all the residents agreed to have a system which could monitor their physical activities. Such system would be really useful for the caregivers in order to know the activity level of the residents and deviations from the normal behavior. From patients’ and caregivers’ point of view, information about the following would be really useful: walking patterns, sleep analysis, real time location, fall detection and physical activity level.\\

All the participants agreed that assistive technologies and AAL solutions can have beneficial effects on quality of life and health. Majority of the residents feel comfortable in using tablet over smart phone and laptop. Big robots and cameras are not preferred by the residents; but they are open for wearable devices. On contrary, the professionals find robots a valuable contributions in smart care homes. \\

Based on the findings of this study it is recommended that a personalized smart home solution which can monitor daily living activities would be really useful for such nursing homes and even for private homes where old people live alone. Such system should be capable of detecting the user’s activity level and thus sends out recommendations to perform some physical activity when the user is inactive for a while. It can also notify the care staff and/or family members about the abnormalities. 

\section*{Acknowledgement}
This work has been funded by the European Union Horizon2020 MSCA ITN ACROSSING project (GA no. $616757$). The authors would like to thank all the participants for their contributions; Esther Veraa, Cindy Wings, Maarten Coolen from Zuyderland, Sittard-Geleen, Netherlands for their valuable inputs and immense support. 
%
\bibliographystyle{splncs}
\bibliography{references}

\begin{thebibliography}{10}

\bibitem{WPA2015}
DESA, U.:
\newblock United nations department of economic and social affairs/population
  division:world population ageing 2015.
\newblock (2015)

\bibitem{kleinberger2007ambient}
Kleinberger, T., Becker, M., Ras, E., Holzinger, A., M{\"u}ller, P.:
\newblock Ambient intelligence in assisted living: enable elderly people to
  handle future interfaces.
\newblock In: International Conference on Universal Access in Human-Computer
  Interaction, Springer (2007)  103--112

\bibitem{olderpeople}
Molinuevo, D.:
\newblock Services for older people in europe.
\newblock (October, 2008)

\bibitem{longtermcare}
Bettio, F., Verashchagina, A.:
\newblock Long-term care for the elderly. provisions and providers in 33
  european countries.
\newblock (November, 2010)

\bibitem{ribbe1997nursing}
Ribbe, M.W., Ljunggren, G., Steel, K., Topinkova, E., Hawes, C., Ikegami, N.,
  Henrard, J.C., J{\'O}Nnson, P.V.:
\newblock Nursing homes in 10 nations: a comparison between countries and
  settings.
\newblock Age and ageing \textbf{26}(suppl 2) (1997)  3--12

\bibitem{blackman2016ambient}
Blackman, S., Matlo, C., Bobrovitskiy, C., Waldoch, A., Fang, M.L., Jackson,
  P., Mihailidis, A., Nyg{\aa}rd, L., Astell, A., Sixsmith, A.:
\newblock Ambient assisted living technologies for aging well: A scoping
  review.
\newblock Journal of Intelligent Systems \textbf{25}(1) (2016)  55--69

\bibitem{rashidi2013survey}
Rashidi, P., Mihailidis, A.:
\newblock A survey on ambient-assisted living tools for older adults.
\newblock IEEE journal of biomedical and health informatics \textbf{17}(3)
  (2013)  579--590

\bibitem{chan2009smart}
Chan, M., Campo, E., Est{\`e}ve, D., Fourniols, J.Y.:
\newblock Smart homes—current features and future perspectives.
\newblock Maturitas \textbf{64}(2) (2009)  90--97

\bibitem{das2002role}
Das, S.K., Cook, D.J., Battacharya, A., Heierman, E.O., Lin, T.Y.:
\newblock The role of prediction algorithms in the mavhome smart home
  architecture.
\newblock IEEE Wireless Communications \textbf{9}(6) (2002)  77--84

\bibitem{helal2005gator}
Helal, S., Mann, W., El-Zabadani, H., King, J., Kaddoura, Y., Jansen, E.:
\newblock The gator tech smart house: A programmable pervasive space.
\newblock Computer \textbf{38}(3) (2005)  50--60

\bibitem{rashidi2008keeping}
Rashidi, P., Cook, D.J.:
\newblock Keeping the intelligent environment resident in the loop.
\newblock (2008)

\bibitem{krumm2000multi}
Krumm, J., Harris, S., Meyers, B., Brumitt, B., Hale, M., Shafer, S.:
\newblock Multi-camera multi-person tracking for easyliving.
\newblock In: Visual Surveillance, 2000. Proceedings. Third IEEE International
  Workshop on, IEEE (2000)  3--10

\bibitem{skubic2009smart}
Skubic, M., Alexander, G., Popescu, M., Rantz, M., Keller, J.:
\newblock A smart home application to eldercare: Current status and lessons
  learned.
\newblock Technology and Health Care \textbf{17}(3) (2009)  183--201

\bibitem{ni2015elderly}
Ni, Q., Garc{\'\i}a~Hernando, A.B., de~la Cruz, I.P.:
\newblock The elderly’s independent living in smart homes: A characterization
  of activities and sensing infrastructure survey to facilitate services
  development.
\newblock Sensors \textbf{15}(5) (2015)  11312--11362

\bibitem{holzinger2005extreme}
Holzinger, A., Errath, M., Searle, G., Thurnher, B., Slany, W.:
\newblock From extreme programming and usability engineering to extreme
  usability in software engineering education (xp+ ue/spl rarr/xu).
\newblock In: Computer Software and Applications Conference, 2005. COMPSAC
  2005. 29th Annual International. Volume~2., IEEE (2005)  169--172

\bibitem{tapia2004activity}
Tapia, E.M., Intille, S.S., Larson, K.:
\newblock Activity recognition in the home using simple and ubiquitous sensors.
\newblock In: International Conference on Pervasive Computing, Springer (2004)
  158--175

\bibitem{nguyen2005learning}
Nguyen, N.T., Phung, D.Q., Venkatesh, S., Bui, H.:
\newblock Learning and detecting activities from movement trajectories using
  the hierarchical hidden markov model.
\newblock In: Computer Vision and Pattern Recognition, 2005. CVPR 2005. IEEE
  Computer Society Conference on. Volume~2., IEEE (2005)  955--960

\bibitem{nazerfard2010conditional}
Nazerfard, E., Das, B., Holder, L.B., Cook, D.J.:
\newblock Conditional random fields for activity recognition in smart
  environments.
\newblock In: Proceedings of the 1st ACM International Health Informatics
  Symposium, ACM (2010)  282--286

\bibitem{sun2009promises}
Sun, H., De~Florio, V., Gui, N., Blondia, C.:
\newblock Promises and challenges of ambient assisted living systems.
\newblock In: Information Technology: New Generations, 2009. ITNG'09. Sixth
  International Conference on, Ieee (2009)  1201--1207

\bibitem{dongen2016successful}
Dongen, J.J.J., Habets, I.G.J., Beurskens, A., Bokhoven, M.A.:
\newblock Successful participation of patients in interprofessional team
  meetings: A qualitative study.
\newblock Health Expectations (2016)

\bibitem{demiris2004older}
Demiris, G., Rantz, M.J., Aud, M.A., Marek, K.D., Tyrer, H.W., Skubic, M.,
  Hussam, A.A.:
\newblock Older adults' attitudes towards and perceptions of ‘smart
  home’technologies: a pilot study.
\newblock Medical informatics and the Internet in medicine \textbf{29}(2)
  (2004)  87--94

\bibitem{alam2012review}
Alam, M.R., Reaz, M.B.I., Ali, M.A.M.:
\newblock A review of smart homes—past, present, and future.
\newblock IEEE Transactions on Systems, Man, and Cybernetics, Part C
  (Applications and Reviews) \textbf{42}(6) (2012)  1190--1203

\bibitem{netuveli2008quality}
Netuveli, G., Blane, D.:
\newblock Quality of life in older ages.
\newblock British medical bulletin \textbf{85}(1) (2008)  113--126

\bibitem{siegel2014contributions}
Siegel, C., Hochgatterer, A., Dorner, T.E.:
\newblock Contributions of ambient assisted living for health and quality of
  life in the elderly and care services-a qualitative analysis from the
  experts’ perspective of care service professionals.
\newblock BMC geriatrics \textbf{14}(1) (2014)  112

\bibitem{holzinger2008investigation}
Holzinger, A., Schaupp, K., Eder-Halbedl, W.:
\newblock An investigation on acceptance of ubiquitous devices for the elderly
  in a geriatric hospital environment: using the example of person tracking.
\newblock In: International Conference on Computers for Handicapped Persons,
  Springer (2008)  22--29

\bibitem{elo2008qualitative}
Elo, S., Kyng{\"a}s, H.:
\newblock The qualitative content analysis process.
\newblock Journal of advanced nursing \textbf{62}(1) (2008)  107--115

\bibitem{miles1994qualitative}
Miles, M.B., Huberman, A.M.:
\newblock Qualitative data analysis: An expanded sourcebook.
\newblock sage (1994)

\bibitem{krefting1991rigor}
Krefting, L.:
\newblock Rigor in qualitative research: The assessment of trustworthiness.
\newblock American journal of occupational therapy \textbf{45}(3) (1991)
  214--222

\end{thebibliography}

\end{document}